\documentclass[12pt, letterpaper, titlepage]{article}
\usepackage{fullpage}
\usepackage[margin=1in, nohead]{geometry}
\usepackage{amsmath}
\usepackage{graphicx}
\usepackage{amsfonts}
\usepackage{amssymb}
\usepackage{titlesec}
\usepackage{enumitem}
\usepackage{amsmath}
\usepackage{setspace}
\usepackage[notablist]{endfloat}
\usepackage{booktabs, caption, makecell}
\usepackage{authblk}
\usepackage{natbib,upgreek}
\usepackage{nicefrac}
\usepackage{textcomp}
\usepackage{siunitx}
\usepackage{multirow}
\usepackage{graphicx}
\usepackage{caption}

\newcommand{\Tj}{\boldsymbol{\tau}_{j}}
\newcommand{\Tji}{\tau_{j,i}}

\newcommand{\Yji}{y(\Tji)}
\newcommand{\bYj}{\mathbf{y}_{\Tj}}
\newcommand{\bY}{\mathbf{y}_{\boldsymbol{\tau}}}
\newcommand{\lj}{\lambda_{j}}
\newcommand{\lbj}{\lambda_{j,b}}
\newcommand{\lfj}{\lambda_{j,f}}
\newcommand{\ljt}{\lambda_{j}(t)}

\newcommand{\lbjt}{\lambda_{j,b}(t)}
\newcommand{\lfjt}{\lambda_{j,f}(t)}
\newcommand{\muTji}{\mu(\Tji)}

\newcommand{\BObj}{\beta_{0,j}^{(b)}}
\newcommand{\BOfj}{\beta_{0,j}^{(f)}}

\newcommand{\BOb}{\boldsymbol{\beta}_{0}^{(b)}}

\newcommand{\Bb}{\boldsymbol{\beta}^{(b)}}

\newcommand{\Bbj}{\boldsymbol{\beta}_{j}^{(b)}}
\newcommand{\Bfj}{\boldsymbol{\beta}_{j}^{(f)}}
\newcommand{\bx}{\mathbf{x}}
\newcommand{\bX}{\mathbf{X}}

\newcommand{\bxj}{\mathbf{x}(\tau_{j,i})}
\newcommand{\tlog}{\text{log}}

\newcommand{\bzero}{\mathbf{0}}

\newcommand{\bSj}{\mathbf{S}_{j}}

\newcommand{\bpB}{\boldsymbol{\sigma}^{2}_{b}}

\newcommand{\pBj}{\sigma^{2}_{b,j}}
\newcommand{\pFj}{\sigma^{2}_{f,j}}
\newcommand{\sigzero}{\sigma_{0}^{2}}
\newcommand{\taub}{\tau_{b}^{2}}

\newcommand{\one}{\mathbf{1}}
\newcommand{\PFtj}{P_{j}(t)}
\newcommand{\PFTj}{P_{j}(\tau_{j,i})}

\newcommand{\bTstar}{\boldsymbol{\tau}^{*}}
\newcommand{\Tstar}{\tau^{*}}

\newcommand{\bA}{\mathbf{A}}
\newcommand{\nstar}{n^{*}}

\newcommand{\pstar}{p^{*}}
\newcommand{\bgmean}{\boldsymbol{\mu}_{b}}
\newcommand{\bgvar}{\boldsymbol{\Sigma}_{b}}
\newcommand{\eye}{\mathbf{I}}
\newcommand{\bR}{\mathbf{R}}
\newcommand{\bH}{\mathbf{H}}

\newcommand{\bL}{\mathbf{L}}
\newcommand{\bSstar}{\mathbf{S}^{*}}
\newcommand{\bSstarinv}{\mathbf{S}^{*-1}}
\newcommand{\bSb}{\mathbf{S}_{b}}

\newcommand{\bth}{\boldsymbol{\theta}}
\newcommand{\bs}{\mathbf{s}}

\newcommand{\Dstar}{D^{*}}
\newcommand{\yho}{y_{\text{ho}}(\tau_{j,i})}
\newcommand{\BOstar}{\beta_{0,j}^{*}}
\newcommand{\bstar}{\boldsymbol{\beta}_{j}^{*}}

\begin{document}

\title{A Model-Based Approach to Wildland Fire Reconstruction Using Sediment Charcoal Records}

\author{M.S. Itter \thanks{ittermal@msu.edu}\\Department of Forestry, Michigan State University, East Lansing, MI\\
\and A.O. Finley\\Department of Forestry, Michigan State University, East Lansing, MI\\
Department of Geography, Michigan State University, East Lansing, MI
\and M.B. Hooten\\U.S. Geological Survey, Colorado Cooperative Fish and Wildlife Research Unit, Fort Collins, CO\\
Department of Fish, Wildlife, and Conservation Biology, Colorado State University, Fort Collins, CO\\
Department of Statistics, Colorado State University, Fort Collins, CO
\and P.E. Higuera\\Department of Ecosystem and Conservation Sciences, University of Montana, Missoula, MT
\and J.R. Marlon\\Yale School of Forestry and Environmental Studies, Yale University, New Haven, CT
\and R. Kelly\\Neptune and Company Inc., Durham, NC
\and J.S. McLachlan\\Department of Biological Sciences, University of Notre Dame, South Bend, IN}

\date{}

\maketitle

\section*{Abstract}
Lake sediment charcoal records are used in paleoecological analyses to reconstruct fire history including the identification of past wildland fires. One challenge of applying sediment charcoal records to infer fire history is the separation of charcoal associated with local fire occurrence and charcoal originating from regional fire activity. Despite a variety of methods to identify local fires from sediment charcoal records, an integrated statistical framework for fire reconstruction is lacking. We develop a Bayesian point process model to estimate probability of fire associated with charcoal counts from individual-lake sediments and estimate mean fire return intervals. A multivariate extension of the model combines records from multiple lakes to reduce uncertainty in local fire identification and estimate a regional mean fire return interval. The univariate and multivariate models are applied to 13 lakes in the Yukon Flats region of Alaska. Both models resulted in similar mean fire return intervals (100-350 years) with reduced uncertainty under the multivariate model 
due to improved estimation of regional charcoal deposition. The point process model offers an integrated statistical framework for paleo-fire reconstruction and extends existing methods to infer regional fire history from multiple lake records with uncertainty following directly from posterior distributions.

\section*{keywords}
paleoecology, fire history, fire return interval, Bayesian hierarchical model, Poisson point process

\section{Introduction}\label{sect:intro}
Charcoal particles deposited in lake sediments during and following wildland fires serve as records of local to regional fire history. Sediment 
charcoal records are used in paleoecological analyses to identify individual fire events and to estimate fire frequency and regional biomass 
burned at centennial to millennial time scales \citep{Clark1988, Clark1990, Whitlock1996, Long1998, Power2008}. When combined 
with sediment pollen records, charcoal deposits can be used to infer relationships between changing climate, vegetation, and fire regimes including 
fire frequency, size, and severity \citep{Clark1996, ClarkRoyall1996, Long1998, Carcaillet2001, Higuera2009, Kelly2013}. In particular, combined 
sediment charcoal records from multiple lakes have been used to correlate changes in regional biomass burned with shifts in regional vegetation and/or climate 
\citep{Power2008, Higuera2009, Marlon2012, Kelly2013}.

Charcoal deposits in lake sediments arise from several different sources. Large charcoal particles ($>\SI{100}{\micro\metre}$) have small dispersal 
distances and exhibit strong correlation with fire occurrence within roughly 500 to 1000 meters of lakes \citep{Clark1988, Whitlock1996, Gavin2003, Lynch2004, Peters2007}. 
Small charcoal particles ($<\SI{50}{\micro\metre}$) have larger dispersal distances (typically 1-20 km) and are indicators of regional biomass burned \citep{Clark1988, 
ClarkRoyall1996}. Sediment charcoal deposits arise from primary sources, direct transport during a fire, as 
well as secondary sources including surface transport via wind and water of charcoal deposited within a lake catchment \citep{Whitlock1996, 
Higuera2007}. Further, lake sediments mix over time, redistributing charcoal particles vertically and concentrating charcoal in the lake center 
\citep{Whitlock1996}. The different depositional sources and sediment mixing increase the variability in sediment charcoal records and make 
inference regarding the size and location of individual fire events difficult \citep{Higuera2007}. Despite the noise present in sediment charcoal records, the use of such data to 
accurately identify local fire events has been consistently demonstrated \citep{Clark1990, Gavin2003, Lynch2004, Higuera2007}.

Charcoal deposition is often expressed in terms of charcoal accumulation rate to account for different sedimentation rates over time (CHAR; particles $\cdot$ cm$^{-2} \cdot$ yr$^{-1}$). 
Analytical approaches to identify individual, local fire events based on sediment charcoal records decompose CHAR into background and peak components. The 
background component captures low-frequency variability associated with time-varying charcoal production rates (e.g., changes in biomass burned), secondary charcoal deposition, 
sediment mixing, and charcoal arising from regional sources. The peak component captures high-frequency variability associated with local fire events 
as well as measurement and random error \citep{ClarkRoyall1996, Long1998}. Approaches to estimate background accumulation include low-pass filters applied to 
Fourier-transformed CHAR \citep{ClarkRoyall1996, Carcaillet2001} and locally-weighted regression models \citep{Long1998, Gavin2006, 
Higuera2009}. Charcoal peaks are defined as the residuals resulting from raw CHAR series minus background CHAR or the ratio between raw and background CHAR. 
A threshold is used to distinguish charcoal peaks indicative of local fire events from false peaks attributable to elevated background deposition \citep{Clark1996}. 
Optimal thresholds, in terms of correct identification of local fires, are estimated using sensitivity analysis \citep{Clark1996} or upper 
quantiles of a Gaussian mixture model \citep{Gavin2006}, lacking independent fire records to identify and validate optimal threshold values \citep{Higuera2009}.

While methods to identify local fire events based on sediment charcoal records have been well developed over the past 30 years, an integrated statistical framework 
for fire identification is still lacking \citep{Higuera2010}. We build upon existing charcoal analysis methods to develop a hierarchical Bayesian Poisson point process model 
for fire identification and estimation of fire return intervals (FRIs). The point process model offers a fully model-based approach to charcoal analysis with several 
important properties. The model operates on charcoal counts directly, using an offset term to control for sedimentation rate. We generate an explicit probability 
of fire estimate for each charcoal count. The hierarchical Bayesian approach makes for tractable error propagation allowing 
for a complete treatment of uncertainty sources in sediment charcoal records including uncertainty associated with sediment age models. The model is easily extended to multivariate data sets, allowing for pooling of 
sediment charcoal records among lakes. While methods currently exist to pool charcoal records \citep{Power2008}, the Poisson point process model requires no 
transformation or interpolation of charcoal counts, improving interpretability of results and avoiding potential introduction of non-quantifiable error to 
charcoal data sets. The modeling approach objectively identifies parameter values controlling the decomposition of sediment charcoal into background and peak 
components via regularization. Most importantly, the hierarchical Bayesian Poisson point process model provides an integrated probabilistic framework to identify 
local fires and estimate FRIs across multiple lake records with explicit uncertainty quantification.

The remainder of this article is organized as follows. In section \ref{sect:model}, we develop a Poisson point process model for charcoal deposition using 
data from a single lake (section \ref{sect:uvmod}) and a regional network of lakes (section \ref{sect:mvmod}). Estimation of local fire probability and mean 
FRIs are described in the context of developing the single-lake model and the multiple-lake model. Implementation of the two models applying Bayesian inference 
is described in section \ref{sect:bayesimp}. We demonstrate the application of the single-lake and multiple-lake models to both simulated data and 
observed data from a regional network of lakes (section \ref{sect:app}). We conclude with a discussion of modeling properties and results (section \ref{sect:disc}).

\section{Bayesian Point Process Model for Charcoal Deposition}\label{sect:model}
We construct a Bayesian Poisson point process model that relates charcoal deposition in lake sediments to local and regional fire occurrence. Charcoal particles 
arising from different sources (i.e., regional and secondary sources versus local fire events) are indistinguishable in sediment charcoal records (apart from the size 
distinction noted earlier). We separate 
background from peak deposition by assuming charcoal particles are generated by independent processes in time: a smooth background process, exhibiting low-frequency changes 
in charcoal deposition rates over time, and a highly variable foreground (or peak) process, exhibiting high-frequency changes in charcoal deposition rates 
associated with local fire events. Total charcoal deposition is proportional to the sum of the background and foreground processes \citep{Clark1996}. The separation of 
total charcoal deposition into background and foreground processes provides the necessary analytical mechanism to identify local fire events from noisy sediment charcoal records. We 
begin this section by defining a univariate Poisson point process model to identify local fires events. We then extend the univariate model to 
accommodate sediment charcoal records from a regional network of lakes using a multivariate model.

\subsection{Univariate Model}\label{sect:uvmod}
Charcoal counts are observed over time intervals spanned by the bottom and top ages of a sediment core section: $\tau_{j,i} = t_{j,i}^{(b)} - t_{j,i}^{(a)}$, where $t_{j,i}^{(b)}$ 
and $t_{j,i}^{(a)}$ are the bottom and top ages of sediment core section $i$ ($i=1,\ldots,n_j$) from lake $j$ ($j=1,\ldots,k$). The $\tau_{j,i}$ correspond to 
non-overlapping time intervals such that $\bigcup_{i=1}^{n_j}\Tji = D_{j}$ where $D_{j}$ is the temporal domain of lake $j$. Throughout, we use $\Tj$ to denote 
the set of observed time intervals, the temporal support, for lake $j$ (i.e., $\Tj = (\tau_{j,1},\tau_{j,2},\ldots,\tau_{j,n_j})'$). Let $y(\tau_{j,i})$ equal the observed charcoal count for $\Tji$ 
defining $\bYj = (y(\tau_{j,1}),y(\tau_{j,2}),\ldots,y(\tau_{j,n_{j}}))'$. We model $\bYj$ using a Poisson distribution conditional on a latent intensity process 
$\lj$ that can be decomposed into additive continuous background and foreground intensity processes: $\lj = \lbj + \lfj$, 
where $\lbj$ and $\lfj$ denote the background and foreground intensities, respectively. Then,
\begin{equation}
 \label{eqtn1:likelihood}
 \bYj\lvert\lj \sim \prod_{i=1}^{n_j}\text{Poisson}(\mu(\tau_{j,i})),
\end{equation}
where $\mu(\tau_{j,i})$ arises as the temporal aggregation of the continuous intensity process. That is,
\[\muTji = \int_{\Tji}\ljt{dt}, \qquad t\in{\Tji},\]
for $i = 1,\ldots,n_{j}$ and $j=1,\ldots,k$. The background intensity is a smooth process capturing charcoal influx from regional, secondary sources and is defined 
by low-frequency changes over time. The foreground intensity is a highly variable process capturing charcoal influx from local fire events and 
is defined by high-frequency changes over time.

We model the continuous background and foreground intensity processes on the log scale for $t\in{\tau_{j,i}}$ as,
\begin{equation}
\label{eqtn2:intmod}
\begin{aligned}
 \tlog\left(\lbj(t)\right) &= \BObj + \bxj'\Bbj\\[5pt]
 \tlog\left(\lfj(t)\right) &= \BOfj + \bxj'\Bfj
\end{aligned}
\end{equation}
where $\BObj$ and $\BOfj$ are intercept terms, $\Bbj$ and $\Bfj$ are $p$-dimensional vectors of regression coefficients, and $\bxj$ is 
a $p$-dimensional set of known covariate values corresponding to basis function values at $p$ knots. We assume the latent intensity processes $\lbj$ and $\lfj$ are independent conditional on their respective 
regression coefficients. For $t \in \tau_{j,i}$, we can express the background and foreground processes at temporal support $\Tj$ as
\begin{equation}
\label{eqtn3:intT}
\begin{aligned}
\lbj(\Tji) &= \int_{\Tji} \lbjt\:dt \; = \; \text{e}^{\BObj + \bxj'\Bbj}\lvert{\Tji}\lvert\\[5pt]
\lfj(\Tji) &= \int_{\Tji} \lfjt\:dt \; = \; \text{e}^{\BOfj + \bxj'\Bfj}\lvert{\Tji}\lvert.
\end{aligned}
\end{equation}

We apply natural cubic splines as our basis functions subject to a penalty term placed on the regression coefficients $\Bbj$ and $\Bfj$, forcing the background process to 
be smooth and the foreground process to be variable. Without the penalty term constraining the background and foreground intensities, the regression coefficients in (\ref{eqtn2:intmod}) 
would be confounded. Alternative spline or predictive process basis functions may also be used. 

Without separating background and foreground processes, charcoal deposition to a lake $j$ represents a continuous, non-homogeneous Poisson process. However, the likelihood for $\bYj$ (\ref{eqtn1:likelihood}), the models for the background and foreground intensity processes defined at 
support $\Tj$ (\ref{eqtn2:intmod}), and the associated model assumptions define a homogeneous Poisson point process for charcoal deposition in lake $j$ with two important properties:
\begin{enumerate}
 \item independent homogeneous Poisson processes exist for each interval $\Tji$ defined by unique values of $\lbj$ and $\lfj$ according to (\ref{eqtn3:intT});
 \item conditional on $\lbj$ and $\lfj$, charcoal particles are deposited independently of one another and with respect to time within and among time intervals.
\end{enumerate}

\subsubsection{Probability of Fire}\label{sect:pfire}
Charcoal influx at time $t$ arising from local fire events is distinguished from regional, secondary sources according to $\PFtj \equiv \text{Pr}\{\text{fire event local to lake }j\text{ at time }t\}$. That is, 
a charcoal particle arriving at lake $j$ at time $t$ generated dependent on $\ljt$ is labeled as a background or foreground particle according to independent Bernoulli trials 
conditional on $\PFtj$ \citep{Diggle2014}. It follows that $\lfjt = \PFtj\ljt$, and $\lbjt = (1-\PFtj)\ljt$ \citep{Ross2010}, so that
\[\PFtj = \frac{\lfj(t)}{\lfj(t) + \lbj(t)}.\]
The probability of fire at temporal support $\Tj$ is estimated as the mean of the continuous probability of fire function over each interval 
$\Tji$. Specifically,
\[\PFTj = \frac{1}{\lvert{\tau_{j,i}}\lvert}\int_{\tau_{j,i}}\frac{\lfj(t)}{\lfj(t)+\lbj(t)}\:dt, \qquad t \in \tau_{j,i},\]
which is equivalent to
\begin{equation}
\label{eqtn4:pfire}
\frac{\lfj(\tau_{j,i})}{\lfj(\tau_{j,i}) + \lbj(\tau_{j,i})}.
\end{equation}
While the coefficients used to estimate the background and foreground charcoal deposition intensities may not be statistically identifiable as expressed in (\ref{eqtn2:intmod}) without additional 
information, the resulting probability of fire as defined in (\ref{eqtn4:pfire}) is identifiable (see Appendix \ref{app:pfire}).

\subsubsection{Mean Fire Return Interval}\label{sect:FRI}
Estimation of mean FRIs requires setting a probability of fire threshold above which an 
observed charcoal count is considered indicative of a local fire and below which charcoal counts are attributed to regional, secondary charcoal sources. 
The series of probability of fire estimates are transformed into binary values with ones indicating local fire events after selecting an appropriate threshold. Specifically,
\[Z(\tau_{j,i}) = \left\{\begin{array}{ll}
                          1 & P_{j}(\tau_{j,i})>\xi,\\
                          0 & P_{j}(\tau_{j,i})\leqslant\xi
                         \end{array}\right., \]
where $\xi$ is the probability of fire threshold.

The series of observed fire events resulting from the application of the probability of fire threshold constitute a temporal Poisson process defined by 
a rate parameter ($\alpha^{-1}$). 
A property of Poisson processes is the interarrival times, in this case the time intervals between local fire events, are independently and identically distributed as exponential random variables with mean 
$\alpha$. The exponential mean represents the mean FRI, while its inverse, the Poisson rate parameter represents the frequency of local fires. The maximum 
likelihood estimate (MLE) for $\alpha$ is equal to the observation period divided by the number of fire events observed for a given lake, for example,
\[\hat{\alpha}_{j} = \frac{\lvert{D_{j}}\lvert}{\sum_{i=1}^{n_j}Z(\tau_{j,i})},\]
where $\lvert{D_{j}}\lvert$ denotes the length of the temporal domain for lake $j$.

We apply Bayesian inference to estimate the exponential mean parameter ($\alpha$) as described in section \ref{sect:bayesimp}, rather than 
calculating the MLE, to allow for estimation of a regional mean FRI (see section \ref{sect:regFRI}). It is common in fire ecology to apply a Weibull likelihood function to estimate 
the mean FRI. Applying a Weibull likelihood function allows for the probability of a fire event to increase as a function of time elapsed since the last fire event. We did 
not choose to apply a Weibull likelihood function in the current analysis given sediment charcoal records are of relatively short length for a number of study lakes, leading to poor estimation of 
the two Weibull likelihood parameters.

\subsection{Multivariate Model}\label{sect:mvmod}
\begin{figure}
 \centering
 \includegraphics[trim = 1in 4.5in 1in 1in, clip=true]{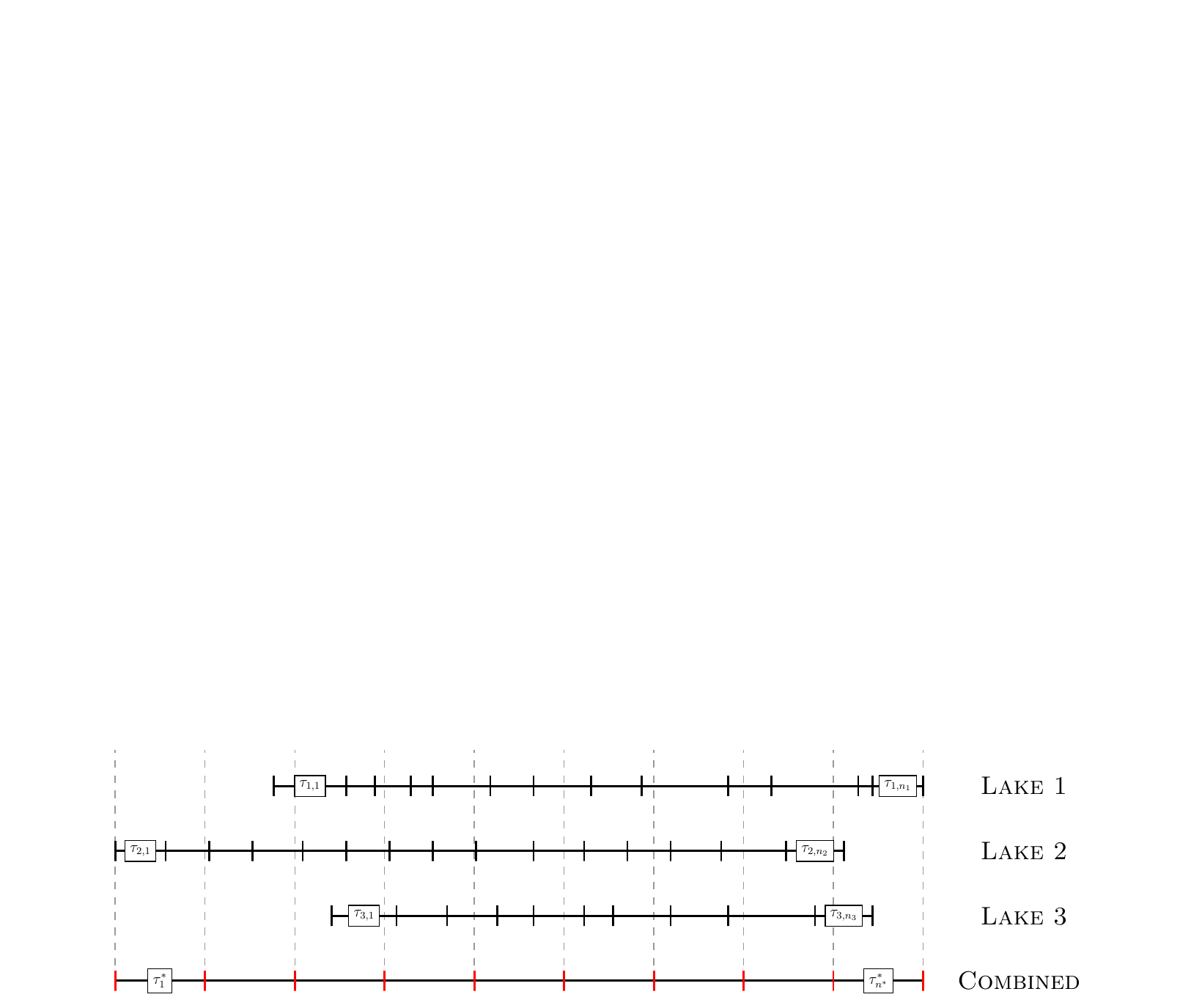}
 \caption[Example creation of new temporal support $\bTstar$ from three individual lake records. Note the temporal misalignment across lakes and $\bTstar$.]{}
 \label{fig:COS}
\end{figure}

The multivariate model follows directly from the univariate model and allows for joint estimation of fire probabilities and mean FRIs across multiple lakes. We combine observations from all lakes $\bY \equiv (\mathbf{y}_{\boldsymbol{\tau}_1}',\mathbf{y}_{\boldsymbol{\tau}_2}'\ldots,\mathbf{y}_{\boldsymbol{\tau}_k}')'$ and 
define new temporal support $\bTstar = \left(\Tstar_{1},\Tstar_{2},\ldots,\Tstar_{\nstar}\right)'$ to accommodate the temporal misalignment among individual lake records 
(Figure \ref{fig:COS}). $\bTstar$ is defined by non-overlapping intervals of equal length 
${\lvert}\Tstar_{i}{\lvert} = {\lvert}\Tstar_{i'}{\lvert}, \; \forall \; i,i' = 1,\ldots,\nstar$, that span $\Dstar$, the temporal domain spanned by all lakes combined 
$\Dstar = \bigcup_{i=1}^{\nstar}\Tstar_{i}$.

The background intensity process for each lake is defined at temporal support $\bTstar$ allowing charcoal counts to be pooled across lakes to estimate 
a regional mean background intensity. Comparable with the univariate model, we model the background intensity process for each 
lake on the log scale, but apply a new set of cubic regression splines defined for $\pstar$ knots corresponding to temporal support $\bTstar$. Specifically, 
for $t\in{\Tstar_{i}}$,
\[\tlog\left(\lbj(t)\right) = \BObj + \bx(\Tstar_{i})'\Bbj,\]
where $\bx(\Tstar_{i})$ is a $\pstar$-dimensional set of known cubic regression spline covariate values equal to the $i$th row of the 
$\nstar \times \pstar$ matrix $\bX$.

The background intensity process defined at temporal support $\bTstar$ is mapped back to the 
observed temporal support for each lake $\boldsymbol{\tau_{j}}$ by an $N \times \nstar$ matrix $\bA$ where $N = \sum_{j=1}^{k} n_j$. 
Specifically, given $\bA \equiv (\bA_{1}',\ldots,\bA_{k}')'$ where each $\bA_{j}$ is an $n_j \times \nstar$ dimensional matrix,
\[\lbj(\tau_{j,i}) = \mathbf{a}_{j,i}'\text{exp}\left(\BObj\one + \bX\Bbj\right)\]
where $\mathbf{a}_{j,i}$ is a $\nstar$-dimensional vector 
equal to the $i$th row of $\bA_{j}$ and $\one$ is a $\nstar$-dimensional vector of ones. The $l$th entry of $\mathbf{a}_{j,i}$ is equal to
\[a_{j,i}(\Tstar_{l}) = \lvert{\tau_{j,i}} \cap \Tstar_{l}\lvert \quad l = 1,\ldots,\nstar\]
such that,
\[\sum_{l=1}^{\nstar} a_{j,i}(\Tstar_{l}) = \lvert{{\tau_{j,i}}}\lvert, \;\: i=(1,\ldots,n_j), \;\; j=(1,\ldots,k).\]
Charcoal counts are pooled across lakes to estimate a regional mean background intensity process by assigning the $\BObj$ and $\Bbj$ exchangeable 
normal prior distributions as described in section \ref{sect:bayesimp}. The foreground intensity process is modeled exactly as in the univariate 
model. Specifically, the foreground process is modeled at the observed temporal support for each lake ($\Tj$) using lake-specific foreground 
coefficients ($\Bfj$) corresponding to a set of $p$ knots. Probability of fire estimates are calculated for each lake independently according to (\ref{eqtn4:pfire}).

\subsubsection{Regional Mean Fire Return Interval}\label{sect:regFRI}
We seek inference regarding the regional mean FRI in addition to individual-lake mean FRIs under the multivariate model. We apply a partial pooling approach 
to estimate the regional mean FRI across lakes. Specifically, individual-lake mean FRI values ($\alpha_{j}$) 
are assigned exchangeable log-normal priors centered on the log of a regional mean FRI ($\text{log}\:\alpha^{*}$) with variance $\sigma_{\text{fri}}^{2}$. 
The partial pooling approach allows charcoal records from each lake to inform the regional average, but penalizes lakes with large uncertainty in their 
mean FRI value estimate. The regional mean FRI variance parameter ($\sigma_{\text{fri}}^{2}$) quantifies the deviation of individual-lake mean FRI 
values from the regional average.

\subsection{Bayesian Implementation}\label{sect:bayesimp}
The univariate and multivariate models are completed by specifying prior distributions for remaining unknown parameters. These include background and 
foreground regression coefficients and mean FRI parameters.

\subsubsection{Univariate Model}\label{sect:imp_uni}
We assigned normal priors to regression coefficients under the univariate model: 
$\BObj \sim \text{N}(0,\sigzero)$, $\Bbj \sim \text{N}(\bzero,\pBj\bSj^{-1})$, $\BOfj \sim \text{N}(0,\sigzero)$, and $\Bfj \sim \text{N}(\bzero,\pFj\bSj^{-1})$. 
Here $\sigzero$ is fixed at a large value defining a diffuse normal prior. The $\pBj$ and $\pFj$ terms represent scalar penalties that are 
regularized subject to the constraint $\pBj < \pFj$ such that the background process is smooth while the foreground process is sufficiently flexible 
to capture irregular charcoal counts (regularization is described in section \ref{sect:reg}). The $p \times p$ matrix $\bSj$ consists of known coefficients defined as a function of the selected knot values \citep{Wood2006}. Note that $\bSj$ is not full column rank, 
rather its rank is $p-2$ given that the second derivative of the boundary knots are equal to zero for natural cubic splines. Thus, the priors for $\Bbj$ and $\Bfj$ are improper, but can be shown to 
result in proper posterior distributions. Combining the likelihood from (\ref{eqtn1:likelihood}) with the priors for the regression parameters, the 
joint posterior distribution for a single lake under the univariate model, using notation similar to \citet{Gelman2014}, is proportional to:
\begin{equation}
 \label{eqtn5:uvpost}
 \prod_{i=1}^{n_j}\text{Pois}(\Yji\lvert{\muTji})\times\text{N}(\BObj\lvert{\sigzero})\times
 \text{N}(\Bbj\lvert{\pBj},\bSj)\times\text{N}(\BOfj\lvert\sigzero)\times\text{N}(\Bfj\lvert\pFj,\bSj).
\end{equation}

\subsubsection{Multivariate Model}\label{sect:imp_mult}
We specified identical normal priors for the foreground regression coefficients under the multivariate model as in the univariate model, and exchangeable 
normal priors for the background coefficients: 
$\BOb \sim \text{N}(\mu_{0,b}\one,\taub\eye_{k})$ and $\Bb \sim \text{N}(\bR\bgmean,\bgvar)$ where $\BOb \equiv \left(\beta_{0,1}^{(b)},\beta_{0,2}^{(b)}\ldots,\beta_{0,k}^{(b)}\right)'$, 
$\Bb \equiv \left(\boldsymbol{\beta}_{1}^{(b)'},\boldsymbol{\beta}_{2}^{(b)'},\ldots,\boldsymbol{\beta}_{k}^{(b)'}\right)'$, 
$\mu_{0,b}$ and $\tau_{b}^{2}$ are univariate mean and variance parameters, $\one$ is a $k$-dimensional vector of ones, 
$\eye_{k}$ denotes a $k$-dimensional identity matrix, $\bgmean$ is a $\pstar$-dimensional mean vector, $\bgvar$ is a $k\pstar \times k\pstar$ covariance 
matrix, and $\bR$ is a $k\pstar \times \pstar$ incidence matrix equal to $\one \otimes \eye_{\pstar}$. The univariate variance $\tau_{b}^{2}$ quantifies 
inter-lake variation in the intercept of the background intensity. The covariance matrix $\bgvar$ 
can be decomposed into inter-lake and within-lake covariance in background coefficients. Defining a $k\pstar \times k\pstar$ block diagonal matrix $\bSb \equiv \text{Diag}(\sigma_{b,1}^{2}\bSstarinv,\sigma_{b,2}^{2}\bSstarinv,\ldots,
\sigma_{b,k}^{2}\bSstarinv)$ where $\bSstar$ is a $\pstar \times \pstar$ matrix of known coefficients associated with the $\pstar$ knots defined for 
$\bTstar$, we can express the background coefficient covariance matrix as $\bgvar = \bL\bSb\bL'$ where $\bL\bL'$ is the Cholesky decomposition 
of $\bH \otimes \eye_{\pstar}$ where $\bH$ is a $k$-dimensional covariance matrix. The matrix $\bSb$ accounts for 
covariance in background coefficients within lakes, while $\bH$ captures covariance among lakes. We apply a spatial covariance 
function to construct $\bH$, although any valid covariance function can be used. Specifically, $\bH \equiv \bH_{s}(\bth_{b})$ where 
$(\bH_{s}(\bth_{b}))_{j,j'} = c(\lvert\lvert\bs_j - \bs_{j'}\lvert\lvert;\bth_{b})$ given $\bs_{j}$ indicates the geographic location of lake $j$ and 
$\bth_{b}$ are unknown spatial covariance parameters.

In the current analysis, we assigned normal priors to the mean intercept $\mu_{0,b} \sim \text{N}\left(0,\sigzero\right)$ and mean regression coefficients 
$\bgmean \sim \text{N}\left(\mathbf{0},\psi^{2}\eye\right)$ where $\sigzero$ is fixed at a large value, while $\psi^{2}$ is set to an appropriate 
order of magnitude for the $\Bbj$ based on univariate model results. The univariate among-lake standard deviation parameter $\tau_{b}$ was assigned a 
uniform prior: $\tau_{b} \sim \text{Unif}(a_{\tau},b_{\tau})$. The scalar penalty on the background deposition process for each lake ($\pBj$) was 
set equal to the regularized background penalty value from the univariate model $\bpB \equiv (\sigma_{b,1}^{2},\sigma_{b,2}^{2},\ldots,\sigma_{b,k}^{2})$. 
We applied an exponential spatial covariance function to form $\bH$: 
$(\bH_{s}(\bth_{b}))_{j,j'} = \sigma_{s}^{2}\text{exp}\left(-\phi\lvert\lvert\bs_j - \bs_{j'}\lvert\lvert\right)$ where $\bth_{b} = (\sigma_{s}^{2},\phi)'$. 
The partial sill ($\sigma_{s}^{2}$) represents spatial variance in background regression coefficients among lakes and has the potential, if it is large, to 
generate highly-variable, unconstrained background deposition processes for individual lakes. To avoid the generation of overly flexible background deposition 
processes, we fixed the partial sill at one ($\sigma_{s}^{2} = 1$). The spatial decay parameter was treated as a free parameter and estimated applying a diffuse 
uniform prior: $\phi \sim \text{Unif}(a_{\phi},b_{\phi})$. Combining the joint likelihood for charcoal counts from all lakes with the priors for the 
multivariate regression model parameters, the joint posterior distribution for all lakes under the multivariate model is proportional to:
\begin{equation}
 \label{eqtn6:mvpost}
 \begin{aligned}
 \prod_{j=1}^{k}\prod_{i=1}^{n_j}\text{Pois}(\Yji\lvert{\muTji})\times\text{N}(\BOb\lvert\mu_{0,b},\taub)\times\text{N}(\Bb\lvert{\bgmean},{\bpB},\bSstar,\phi)\times&\\
 \prod_{j=1}^{k}\text{N}(\BOfj\lvert\sigzero)\times\prod_{j=1}^{k}\text{N}(\Bfj\lvert\pFj,\bSj)
 \times\text{N}(\mu_{0,b}\lvert{\sigzero})\times&\\
 \text{N}(\bgmean\lvert{\psi^{2}})\times\text{Unif}(\tau_{b}\lvert{a_{\tau},b_{\tau}})\times\text{Unif}(\phi\lvert{a_{\phi},b_{\phi}}).&
 \end{aligned}
\end{equation}

We use a Metropolis-within-Gibbs MCMC algorithm \citep{Robert2004} to sample from the posterior distributions in (\ref{eqtn5:uvpost}) and (\ref{eqtn6:mvpost}).

\subsubsection{Regularization}\label{sect:reg}
The background ($\pBj$) and foreground ($\pFj$) penalties are regularized to identify optimal values in the univariate model. 
In the current analysis, we conducted a gridded search over a range of penalty values based on initial exploratory modeling with 
five unique penalty values assigned to the background and foreground: 25 total penalty combinations. The gridded search was conducted applying the univariate point 
process model to a single validation data set for each lake with 25 percent of observations held out. Prediction of held-out charcoal counts was 
carried out via composition sampling using posterior samples of background and foreground coefficients ($\BObj,\Bbj,\BOfj,\Bfj$) to generate 
$\lbj(\Tji)$ and $\lfj(\Tji)$. We sampled $\yho \sim \text{Pois}(\mu(\Tji))$ in a one-for-one fashion, where $\yho$ indicates a held-out 
observation. The resulting samples of $\yho$ represent the posterior predictive distribution of $\yho$. Optimal penalty terms were identified as the background and foreground 
penalties that minimized the posterior predictive loss calculated for the hold-out data \citep{Gelfand1998}. The 
posterior predictive loss rewards accuracy of predictions with a penalty for large variance in predictions indicative of over parameterization. For the multivariate model, 
we applied the optimal background and foreground penalty values identified for each lake under the univariate model.

\subsubsection{Mean FRI}
The mean FRI under the univariate and multivariate models is estimated using FRIs calculated after applying a probability of fire threshold 
to sampled posterior probability of fire values at each iteration of the Gibbs sampler. Specifically, for the $\ell$th iteration of the Gibbs sampler, we obtain a set of 
fire event times $\left(t_{j,1}^{(\text{fire})}, t_{j,2}^{(\text{fire})},\ldots, t_{j,m_j}^{(\text{fire})}\right)^{(\ell)}$, where $m_j$ is the total number of fire events observed for lake $j$, by 
conditioning samples on $Z(\tau_{j,i})^{(\ell)}=1$. A given FRI is equal to the elapsed time between two consecutive fire events. For example, for lake $j$ and iteration $\ell$, the $r$th FRI is given by $\text{FRI}_{j,r}^{(\ell)} = \left(t_{j,r+1}^{(\text{fire})} - t_{j,r}^{(\text{fire})}\right)^{(\ell)}$, for $r = (1,2,\ldots,m_{j}-1)$. 

We assigned a semi-informative, conjugate inverse-gamma prior to the mean FRI parameter $\alpha_{j} \sim \text{InvGamma}(a_{\alpha},b_{\alpha})$ under the univariate model centering its density over 
the possible range of FRIs for study lakes based on previous analyses \citep{Kelly2013}. Combining the prior with the exponential likelihood of the FRIs (see section \ref{sect:FRI}), we obtain an inverse-gamma posterior distribution for $\alpha_{j}$ conditional on the derived set of FRIs. Specifically, for the $\ell$th iteration of the Gibbs sampler, $\textbf{FRI}_{j}^{(\ell)} = \left(\text{FRI}_{j,1},\text{FRI}_{j,2},\ldots,\text{FRI}_{j,m_{j}-1}\right)'$ and $\alpha_{j}^{(\ell)}\lvert\textbf{FRI}_{j}^{(\ell)} \sim \text{InvGamma}(a_{\alpha}+m_{j}^{(\ell)}-1, b_{\alpha}+\sum_{r=1}^{m_{j}^{(\ell)}-1}\text{FRI}_{j,r}^{(\ell)})$.

We seek inference regarding the regional mean FRI in addition to individual-lake mean FRIs under the multivariate model. We assigned diffuse uniform 
priors to the regional mean FRI ($\alpha^{*}$) and the inter-lake standard deviation ($\sigma_{\text{fri}}$). Combining the priors and the exponential likelihood 
of the FRIs, the joint posterior distribution conditional on the model parameters defined in (\ref{eqtn6:mvpost}) is proportional to:
\begin{equation*}
  \prod_{j=1}^{k}\prod_{r=1}^{m_{j}-1}\text{Exp}(\text{FRI}_{j,r}\lvert\alpha_{j}) 
  \times \prod_{j=1}^{k}\text{N}(\text{log}\:\alpha_{j}\lvert\text{log}\:\alpha^{*},\sigma_{\text{fri}}^{2}) 
  \times \text{Unif}(\alpha^{*}\lvert{a_{*},b_{*}}) \times \text{Unif}(\sigma_{\text{fri}}\lvert{a_{\sigma},b_{\sigma}}).
\end{equation*}

We estimated mean FRI values under the univariate and multivariate models using a range of probability of fire thresholds: $\boldsymbol{\xi} = (0.50,0.55,\ldots,1.00)'$. 
Ideally, we would have applied a probability of fire threshold that yielded the greatest accuracy in terms of identifying local fire events (i.e., the most-accurate mean FRI estimate). A common challenge with the use of sediment charcoal records, however, is the lack of independent fire history data to conduct model validation. In the absence of independent fire history data, we sought 
an optimal probability of fire threshold in the sense of providing a precise mean FRI estimate (i.e., similar set of fires identified for each posterior sample of fire 
probabilities). We estimated the coefficient of variation for posterior mean FRI samples ($\nicefrac{\tilde{\sigma}}{\tilde{\mu}}$ where $\tilde{\sigma}$ and $\tilde{\mu}$ are the 
posterior sample standard deviation and mean, respectively). We selected the probability of fire threshold that minimized the coefficient of variation as the optimal threshold. The 
optimal threshold provided the most-consistent mean FRI estimate based on posterior samples.

\section{Model Application}\label{sect:app}
We apply the Poisson point process model to both simulated data and to sediment charcoal records from a 13-lake network in interior 
Alaska. The use of simulated data allows for proper model validation, which is difficult using sediment charcoal records due to the 
lack of observed fire data to compare with probability of fire estimates.

\subsection{Simulation Study}\label{sect:simdat}
We tested the accuracy of the point process model by applying it to simulated sediment charcoal records generated using CharSim \citep{Higuera2007}. 
CharSim is a semi-mechanistic model that generates fires on a landscape and maps the subsequent deposition of charcoal particles to a target lake. 
The amount of charcoal deposited in the target lake is proportional to the size of the fire, its proximity to the target lake, the 
atmospheric injection height of charcoal particles during the fire, and secondary charcoal deposition and sediment mixing within the lake 
\citep[see][for additional details]{Higuera2007}.

We applied the univariate point process model to a simulated sediment charcoal record generated by CharSim to mimic a fire regime consistent with historic fire 
regimes in the Alaskan boreal forest \citep[as described in Table 2 of][]{Higuera2007}.  
We defined a binary variable for each sample interval of the simulated record with one indicating a local fire event within 100 m of the target lake within an interval, 
and 0 indicating no local fires within an interval. Probability of fire estimates from the point process model were converted to binary fire occurrences 
as described in section \ref{sect:FRI} and compared with true fire occurrence values to determine the percentage of fires correctly identified.

\begin{figure}
 \centering
 \includegraphics[trim = 1in 3.45in 1in 3.45in, clip=true]{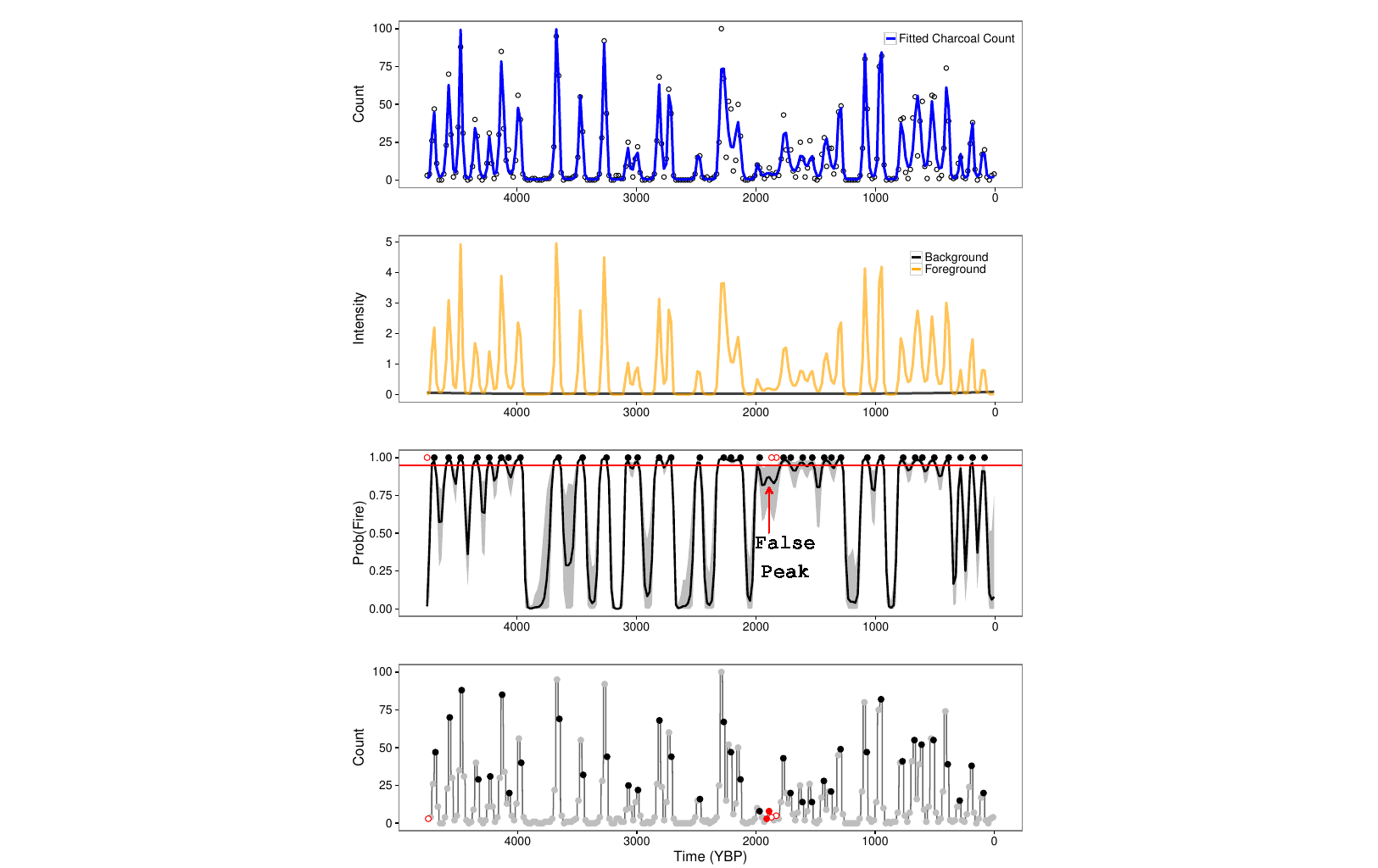}
 \caption[Univariate model results for simulated sediment charcoal record generated by CharSim. Upper panel indicates observed charcoal counts along with 
 the posterior mean charcoal count (blue line). Second panel illustrates posterior mean foreground (orange line) and background (black line) intensities. 
 Third panel plots posterior mean probability of fire estimates for each observed time interval (black line) along with the upper and lower bounds of the 
 95 percent credible interval (gray shading) and the optimal threshold (red line). The points in the third panel correspond to true fire 
 events occurring during the sample interval with black dots delineating correctly identified fires and red open dots delineating missed fires. The arrow highlights 
 the single falsely identified fire during the simulated study period. Lower panel indicates observed charcoal counts with the color and shape indicating whether the count 
 was correctly identified as a true fire (black points), no fire (gray points), missed true fire (red open circles), or falsely identified fire (red points).]{}
 \label{fig:charsim}
\end{figure}

The simulated record was 4760 years long divided into 238 equal-length sample intervals (each interval was 20 years). There were 41 fires within 100 m of the 
target lake leading to a true FRI of 116 years. The point process model correctly identified 38 of the 41 fires occurring over the simulated study period applying an optimal 
threshold value of 0.95, corresponding to a 93 percent fire identification rate (Figure \ref{fig:charsim}). The model was accurate in 
its identification of local fires with only a single falsely identified fire between 1880 to 1920 years before present (YBP). Despite the accuracy of the point process model in 
identifying true local fires, the estimated mean FRI was too long---posterior mean FRI equaled 191 years (95 percent credible interval: 135 to 271 years). The upward bias of the 
point process model in estimating the mean FRI was due to the model's inability to separate fires occurring in close temporal proximity as unique fire events. For example, there were 
40 sample intervals that included a true fire in the simulated record (note, two fires occurred within a single sample interval), but only 25 unique threshold exceedances (or peaks) 
were estimated. A more in-depth discussion of source of bias is provided in section \ref{sect:disc}. Model results were sensitive to the definition of a local fire event. Specifically, 
if a local fire is defined as occurring within 1000 m of the target lake (rather than 100 m), the fire identification rate drops to 75 percent.

\subsection{Yukon Flats}\label{sect:yfdat}
\begin{figure}
 \centering
 \includegraphics[trim = 1in 3.45in 1in 3in, clip=true]{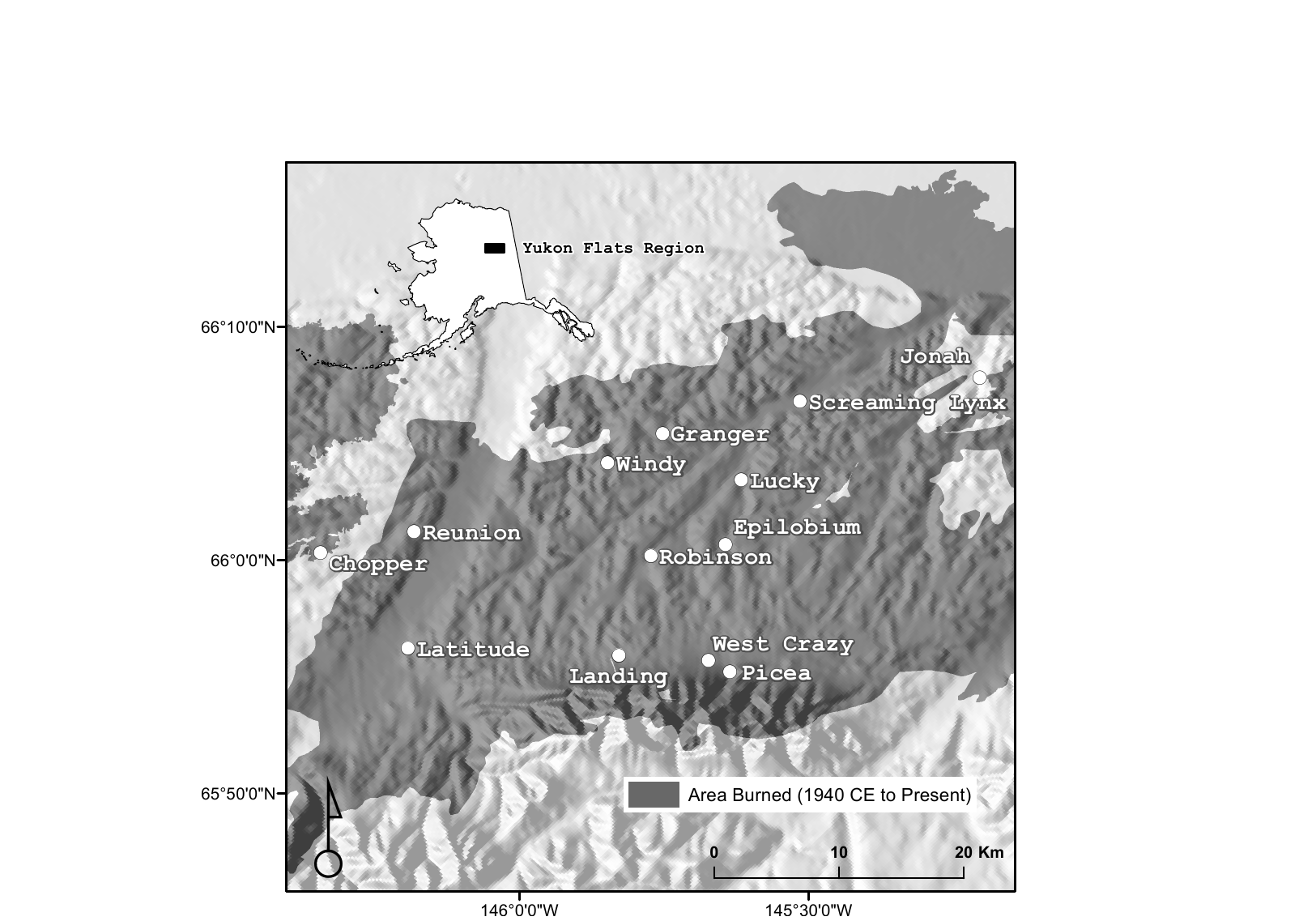}
 \caption[Location of study lakes within the Yukon Flats region of Alaska relative to areas burned in wildland fire since 1940 Common Era.]{}
 \label{fig:map}
\end{figure}

We applied the univariate and multivariate point process models to previously-published sediment charcoal records from 13 lakes in the Yukon Flats region of Alaska 
\citep[Figure \ref{fig:map};][]{Kelly2013}. The Yukon Flats region is dominated by boreal forests and has a fire regime characterized by stand-replacing fires with 
return intervals of several decades to centuries.

\subsubsection{Univariate Model}
\begin{figure}
\centering
\includegraphics[trim = 1in 4.25in 1in 4in, clip=true]{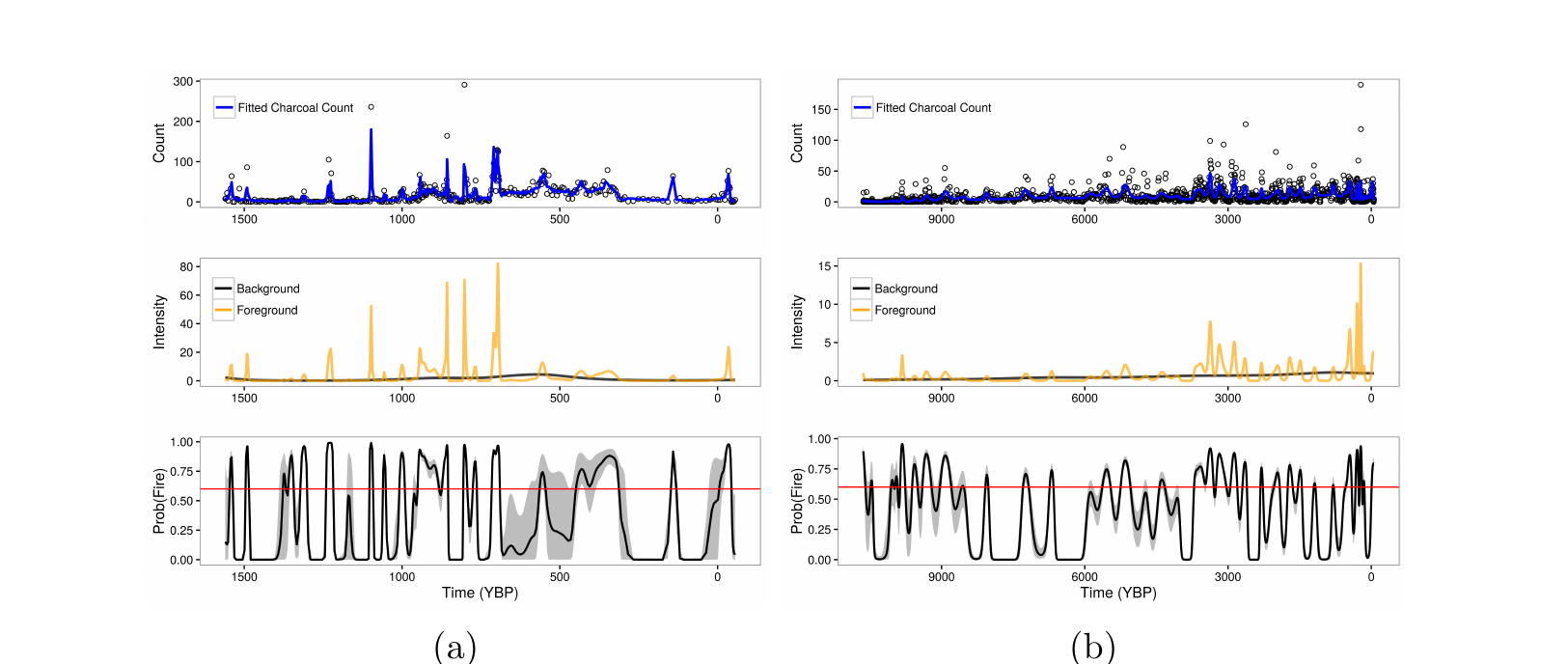}
 \caption[Univariate model results for Chopper Lake (a) and Screaming Lynx Lake (b). Upper panel indicates observed charcoal counts along with 
 the posterior mean charcoal count (blue line). Middle panel illustrates posterior mean foreground (orange line) and background (black line) intensities. 
 Lower panel plots posterior mean probability of fire estimates for each observed time interval (black line) along with the upper and lower bounds of the 
 95 percent credible interval (gray shading) and the optimal threshold (red line).]{}
 \label{fig:uni}
\end{figure}

We applied the univariate model to each of the 13 lakes in the Yukon Flats data set and the multivariate model to all lakes jointly. Mean FRI values 
from the univariate model varied from roughly 134 years for fires local to Chopper Lake to roughly 356 years for fires local to Screaming Lynx Lake. Table 
\ref{tab:fri} provides mean FRI value estimates for each of the 13 lakes. Optimal threshold values providing the most-precise mean FRI estimate 
for each lake varied from 0.50 to 0.75 (Table \ref{tab:fri}). Results of the univariate model for Chopper and Screaming Lynx Lakes are provided in Figure \ref{fig:uni} (similar plots are provided for all lakes in the web supplement; Chopper and Screaming Lynx Lakes are selected to illustrate the lakes with the 
shortest and longest mean FRI, respectively).

Geospatial fire perimeter data for the state of Alaska date back to 1940 Common Era \citep[CE;][]{Alaska2016}. Restricting the fire perimeter data to fires local to lakes within 
the Yukon Flats data set, there were five fires since 1940 within 100 m of one or more of the 13 study lakes (a threshold of 100 m from lake edge was used 
to distinguish local fire events based on results of the simulation study) the earliest of which occurred in 1985. Although the local fire record is insufficient in length 
to conduct proper model validation (i.e., roughly 70 years of local fire data for a study period of over 10,000 years is less than 1 percent data coverage), we can 
compare the true occurrence of local fires to local fires identified by the point process model to assess model performance. Two of the five fires local 
to at least one study lake occurred after the end of the sediment charcoal records for local lakes: Big Creek Fire in 2009, Discovery Creek Fire in 2013. The 
Preacher Creek Fire in 2004 occurred local to Picea and Epilobium Lakes, however, the point process model did not identify a local fire for either of these lakes in the 
most-recent 50 years. There were two unnamed fires, the first in 1985 and the second in 1988, local to several study lakes. A fire event was identified by the point process 
model within the past 25 years in four out of seven lakes local to the 1985 fire and two of five lakes local to the 1988 fire. 

\begin{table}[ht]
\centering
\caption{Summary of univariate and multivariate Poisson process model results for each lake in the Yukon Flats data set. Mean FRI is equal to the posterior mean 
fire return interval with 95 percent credible intervals in parentheses.}
\begin{tabular}{lccc|ccc}
  \hline
\multirow{3}{*}{Lake} & \multicolumn{3}{c|}{Univariate Model} & \multicolumn{3}{c}{Multivariate Model}\\ \cline{2-7}
 & Optimal & Mean & Cred. Int. & Optimal & Mean & Cred. Int.\\
& Threshold & FRI & Width & Threshold & FRI & Width\\
  \hline
Chopper & 0.60 & 134 (91,197) & 106 & 0.85 & 144 (97,216) & 119 \\ 
  Epilobium & 0.50 & 197 (133,292) & 159 & 0.50 & 201 (136,293) & 157 \\ 
  Granger & 0.55 & 286 (187,436) & 249 & 0.50 & 246 (165,368) & 203 \\ 
  Jonah & 0.50 & 159 (111,228) & 117 & 0.55 & 148 (107,207) & 100 \\ 
  Landing & 0.70 & 303 (206,445) & 239 & 0.70 & 291 (201,419) & 218 \\ 
  Latitude & 0.75 & 158 (106,235) & 129 & 0.75 & 137 (96,194) & 98 \\ 
  Lucky & 0.75 & 136 (92,200) & 108 & 0.75 & 138 (93,206) & 114 \\ 
  Picea & 0.50 & 318 (227,448) & 221 & 0.50 & 312 (223,429) & 206 \\ 
  Reunion & 0.70 & 244 (167,354) & 186 & 0.70 & 237 (165,345) & 180 \\ 
  Robinson & 0.55 & 142 (94,213) & 119 & 0.80 & 124 (87,180) & 93 \\ 
  Screaming Lynx & 0.60 & 356 (255,497) & 242 & 0.55 & 357 (255,502) & 247 \\ 
  West Crazy & 0.50 & 204 (133,315) & 182 & 0.70 & 200 (129,306) & 177 \\ 
  Windy & 0.50 & 162 (110,239) & 128 & 0.65 & 155 (107,226) & 119 \\ 
   \hline
\end{tabular}
\label{tab:fri}
\end{table}

\subsubsection{Multivariate Model}
\begin{figure*}
\centering
\includegraphics[trim = 1in 4.25in 1in 4in, clip=true]{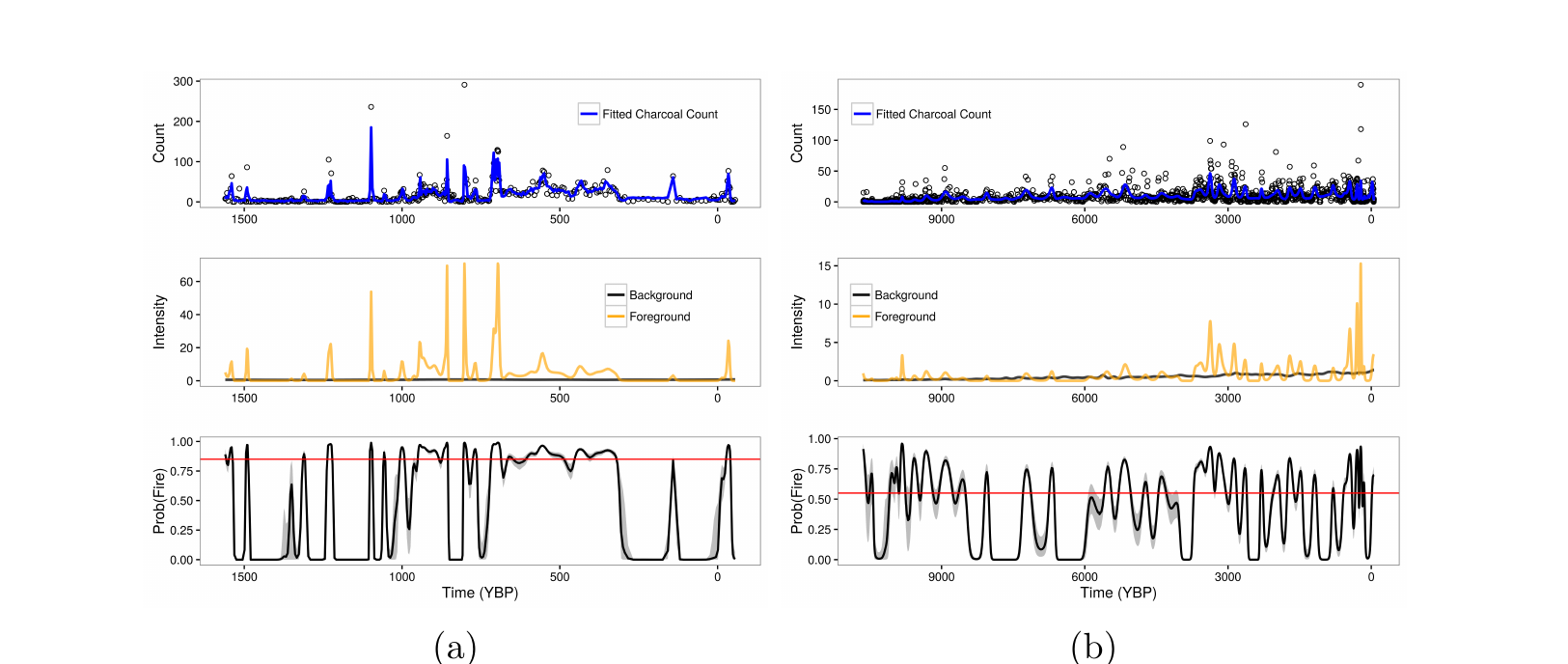}
 \caption[Multivariate model results for Chopper Lake (a) and Screaming Lynx Lake (b). Upper panel indicates observed charcoal counts along with 
 the posterior mean charcoal count (blue line). Middle panel illustrates posterior mean foreground (orange line) and background (black line) intensities. 
 Lower panel plots posterior mean probability of fire estimates for each observed time interval (black line) along with the upper and lower bounds of the 
 95 percent credible interval (gray shading) and the optimal threshold (red line).]{}
 \label{fig:mult}
\end{figure*}

Joint probability of fire estimates generated using the multivariate model varied in magnitude from the univariate model results. Figure \ref{fig:mult} presents 
the multivariate model results for Chopper and Screaming Lynx Lakes. Comparing the results presented in Figure \ref{fig:mult} to those from the univariate model (Figure \ref{fig:uni}), 
the probability of fire estimates for Chopper Lake are slightly higher under the multivariate model than the univariate model, while the probability of fire estimates 
for Screaming Lynx Lake are roughly consistent between the two models. In general, probability of fire estimates were higher under the multivariate model than the univariate model. 
The different magnitudes of probability of fire estimates under the multivariate model necessitated 
calculating new optimal fire thresholds for each lake (Table \ref{tab:fri}). Optimal threshold values ranged from 0.50 to 0.85 for the multivariate model consistent with slightly higher 
probability of fire estimates compared to the univariate model. Mean FRI values 
estimated using the multivariate model were consistent with the mean FRI values estimated for each lake using the univariate model; 
however, the credible interval widths for the multivariate model were narrower than under the univariate model. Specifically, 10 out of 13 lakes had narrower credible intervals under the multivariate model than the 
univariate model (Table \ref{tab:fri}). The average credible interval width for the mean FRI was 156 years under the multivariate model versus 168 years under the univariate model. In addition to 
lake-specific, local mean FRIs, we applied the joint probability of fire estimates and optimal thresholds for each lake to estimate a joint regional mean FRI as 
described in section \ref{sect:regFRI}. We estimated a regional mean FRI over the study period (10,680 to -59 YBP, relative to 1950 CE) of roughly 187 years (95 percent credible interval: 136 to 261 years) for the Yukon Flats region based on the lake network data.

\begin{figure}
 \centering
\includegraphics[trim = 2in 4.35in 2in 4in, clip=true]{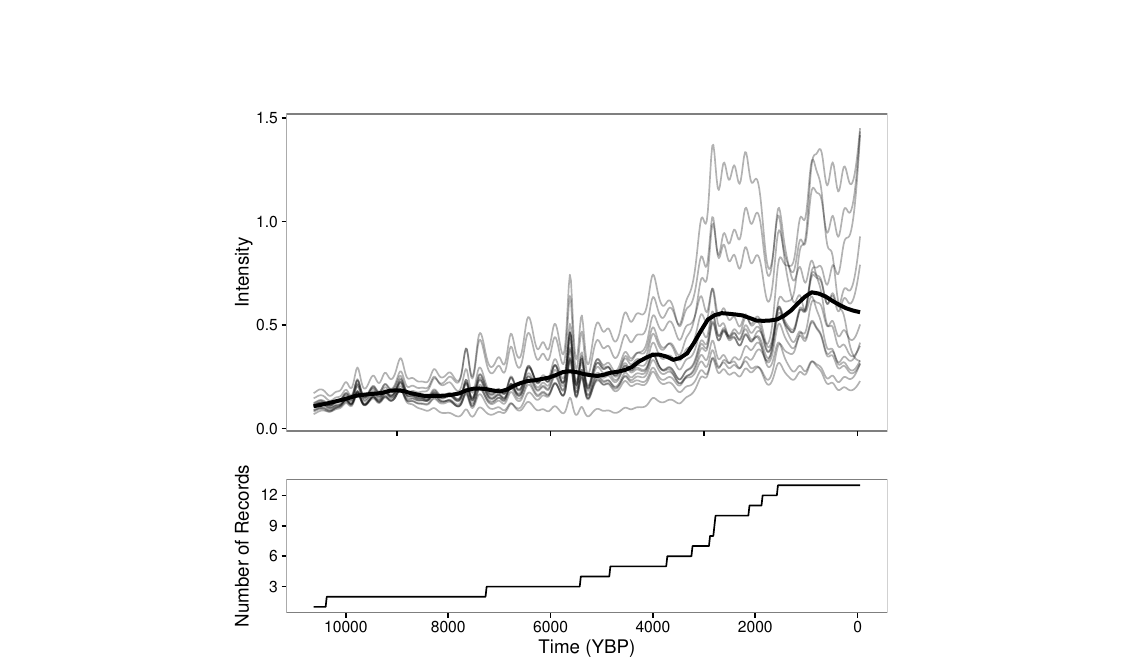}
 \caption[Background charcoal deposition intensity for each lake in the Yukon Flats data set over the entire study period (10,680 to -59 YBP, relative to 1950 CE) based on the multivariate point process model. The bold line in the upper panel is a regional loess smooth function reflecting mean changes in background charcoal deposition across all lakes (fit using a span of 0.15). The lower panel indicates the number of lake records used to estimate background charcoal deposition over time.]{}
 \label{fig:jointBg}
\end{figure}

The multivariate model provides inference on regional background charcoal deposition. Applying an exponential spatial covariance function to describe spatial correlation in 
background regression coefficients (as described in section \ref{sect:imp_mult}) resulted in an estimated effective spatial range of approximately 16.5 km 
(95 percent credible interval: 14.4 to 19.1 km) where the effective spatial range is defined as the distance at which the correlation drops to 0.05. This suggests the parameters describing local, background charcoal intensity for individual lakes are similar for lakes within 20 km of each other. 
Finally, the multivariate model provides an estimate of the background charcoal deposition in each lake over the entire study period, 
although the sediment charcoal records for most lakes are shorter than the full study period. The background charcoal intensities for each lake in the Yukon Flats network are plotted together in Figure  
\ref{fig:jointBg} along with a regional loess smooth function. The background charcoal deposition for most lakes exhibited a similar pattern, with a long-term increase in background charcoal deposition from 6000 YBP to present, a sharp 
increase in background deposition roughly 3000 YBP, and a secondary increase 1000 YBP followed by a decrease in background deposition roughly 500 YBP.

\section{Discussion}\label{sect:disc}
The use of sediment charcoal records to reconstruct past fire regimes is challenging given charcoal counts rather than past fire occurrences are observed. Further, observed 
charcoal counts include charcoal generated during local fires as well as charcoal stemming from regional fire activity and secondary sources. The goal of our analysis was to 
construct an integrated statistical framework for local fire identification and the estimation of mean FRIs based on sediment charcoal records from individual 
lakes building on previous approaches to paleo-fire reconstruction. We further sought to advance existing approaches to reconstruct regional fire history through the development of 
a multivariate model, which combines sediment charcoal records from multiple lakes to identify local fires and jointly estimate mean FRIs at individual-lake and regional scales. Here, we discuss the key results of the application of the univariate and multivariate point process models to the Yukon Flats data set and connect the results of the current analysis to 
previous studies in the same region.

Mean FRI estimates from the univariate and multivariate point process models applied to the Yukon Flats data set ranged from 100 to 350 years (Table \ref{tab:fri}). The lakes with the largest mean FRI estimates have sediment charcoal records dating back the longest among study lakes: Reunion, Granger, Landing, Picea, and Screaming Lynx Lakes all have records that date back at least 5000 YBP. This pattern in consistent with previous interpretations of Holocene fire history in Alaskan boreal forests, which highlight increased fire activity over the last several thousand 
years beginning with the local arrival of black spruce between 6000 to 3000 YBP \citep{Lynch2004, Higuera2009, Kelly2013}.
We observe a similar pattern of increased fire activity in plots of the background charcoal deposition intensity for each lake over the study period derived from the 
multivariate model (Figure \ref{fig:jointBg}). The increase in fires across the Yukon Flats region from 6000 to 3000 YBP reflected in the background intensity suggests that lakes with sediment charcoal records dating back prior to 3000 YBP should have longer mean FRI values than lakes with relatively short records. A secondary peak in the background charcoal 
deposition intensity is observable around 500 YBP coincident with the Medieval Climate Anomaly, a period of increased temperatures and drought frequency (1000-500 YBP), followed by 
the Little Ice Age, a period of cooler and wetter climatic conditions (500-80 YBP). Finally, modeled background charcoal intensities for individual lakes indicate an increase in biomass burned in recent decades, although the increase is not reflected in the regional loess smoother (Figure \ref{fig:jointBg}). The modeled background charcoal intensities are consistent with composite 
CHAR records (i.e., mean charcoal accumulation rate among lakes) calculated using the Yukon Flats data \citep{Kelly2013}.

The univariate Bayesian Poisson point process model provides a model-based approach to estimate probability of fire values associated with sediment charcoal records from a single lake and convert those probabilities into mean FRI estimates. The multivariate model allows for correlation among lakes in the parameters used to estimate the background charcoal deposition intensity. 
The background intensity reflects regional charcoal sources and exhibits low-frequency changes over time associated with factors such as species composition and climate. As such, background charcoal deposition should be similar among lakes in the same region with a high potential for correlation in background deposition process parameters. As expected, mean FRI estimates for each lake are similar based on the univariate and multivariate models (Table \ref{tab:fri}). The multivariate model, however, resulted in mean FRI estimates with reduced uncertainty. Specifically, the 95 percent credible interval for the mean FRI was narrower in 10 out of 13 lakes in the Yukon Flats network with a mean credible interval width of 156 years for the multivariate model versus 168 years for the univariate model. The reduced uncertainty in mean FRI estimates 
under the multivariate model provides some evidence that background charcoal deposition is indeed correlated among lakes located in the same region and that we can reduce uncertainty in estimates of background 
charcoal deposition by accounting for such correlation. The effective spatial range for the background process regression parameters ($\Bbj$) was estimated to be 16.5 km and provides some indication of the distance within which background charcoal deposition is similar among lakes within the Yukon Flats region. This estimate is consistent with the previous analysis using the Yukon Flats data set, which found significant correlation between composite CHAR and regional area burned within a 20-km radius \citep{Kelly2013}.

We estimated a regional mean FRI of roughly 187 years (95 percent credible interval: 136 to 261 years) for the Yukon Flats over the study period applying the partial pooling approach described in section \ref{sect:regFRI}. The partial 
pooling approach also produces estimates of mean FRI values for individual lakes similar to the univariate and multivariate results presented in Table \ref{tab:fri}. However, we do not see the same reduction in uncertainty in individual-lake mean FRI estimates when conducting partial pooling. Specifically, credible interval widths were narrower in only 6 out of 13 lakes with the remaining intervals comparable to univariate 
model results. The partial pooling approach adds two additional parameters ($\alpha^{*},\sigma_{\text{fri}}^{2}$) and combines uncertainty in mean FRI values across lakes. As such, it is not surprising the partial pooling approach does not lead to the same reductions in uncertainty as generating individual-lake mean FRI estimates based on the multivariate model. We envision the partial pooling approach 
being applied only in the setting where a researcher is interested in estimating a regional mean FRI, otherwise, the individual-lake approach is preferred.

The univariate point process model we developed achieves our goal of developing an integrated statistical framework for local fire identification and estimation 
of mean FRIs based on sediment charcoal records for individual lakes. The Bayesian hierarchical model structure allows for tractable propagation of additional 
uncertainty sources in paleo-fire reconstructions. In particular, uncertainty in sediment age models can be integrated by treating the ages of sediment core sections 
as unobserved, latent variables in the point process model. The multivariate extension of the point process model 
provides a novel approach for paleo-fire reconstruction applying multiple lake records to make inferences at both individual-lake and regional scales. Specifically, the 
multivariate model provides estimates of individual-lake mean FRIs, a regional mean FRI, and background charcoal deposition intensity indicative of regional biomass burned. 
When applied to the Yukon Flats data set, pooling of individual-lake records under the multivariate model led to reduced uncertainty in individual-lake mean FRIs. We expect the multivariate model to provide even greater reductions in uncertainty in individual-lake mean FRI values, 
and improved estimates of regional parameters including the regional mean FRI, when applied to larger regional lake networks (i.e., networks with 20 plus lakes), assuming 
all lakes share a common regional fire regime.

Additional information and supporting material for this article is available online at the journal's website.

\section{Acknowledgments}
The authors thank collaborators on the PalEON project and participants of the Paleo-Fire Workshop, Harvard Forest, Sep. 27 to Oct. 2, 2015, Patrick Bartlein in particular, for their useful 
comments and feedback on the modeling approach. This work was supported by National Science Foundation Grants: DMS-1513481, EF-1137309, EF-1241874 and EF-1253225 (M.S. Itter and A.O. Finley); 
MSB-1241856 (M.B. Hooten); and BCS-1437074 (J.R. Marlon). Any use of trade, firm, or product names is for descriptive purposes only and does not imply endorsement by the U.S. Government.

\bibliography{CharcoalModel}
\bibliographystyle{apalike}

\section*{Appendix}

\subsection*{A. Probability of Fire Identification}\label{app:pfire}
The probability of fire for a given sample interval $\tau_{j,i}$ as defined in (\ref{eqtn4:pfire}) can be expressed as
\begin{equation*}
 \frac{\text{e}^{\BOfj + \bxj'\Bfj}}{\text{e}^{\BOfj + \bxj'\Bfj} + \text{e}^{\BObj + \bxj'\Bbj}}
\end{equation*}
which can be simplified to
\begin{equation*}
 \frac{1}{1 + \text{e}^{-\left(\BOstar + \bxj'\bstar\right)}}
\end{equation*}
where $\BOstar = \BOfj - \BObj$ and $\bstar = \Bfj - \Bbj$, thereby proving the identifiability of the probability of fire $\PFTj$. The simplified probability of 
fire is equivalent to the mean response of a logistic regression model fit using a binary variable indicating the occurrence of a local fire within a given sample interval. Specifically, 
we can express the odds of a local fire in a given sample interval as
\begin{equation*}
 \frac{\lfj(\Tji)}{\lbj(\Tji)} = \frac{\text{e}^{\BOfj + \bxj'\Bfj}}{\text{e}^{\BObj + \bxj'\Bbj}}
\end{equation*}
such that the log-odds are given by: $\BOstar + \bxj'\bstar$. Thus, there is a direct connection between the mean probability of fire function defined using Poisson count data in (\ref{eqtn4:pfire}) 
and a logistic regression model used to estimate the probability of fire using Bernoulli observations of fire occurrence/non-occurrence.

\label{lastpage}

\end{document}